\documentclass{moriond}

\def\Journal#1#2#3#4{{#1} {\bf #2}, #3 (#4)}

\def\PRD{{ Phys. Rev.} D}

\def\JCAP{{ JCAP}}
\def\MNRAS{{ MNRAS}}
\def\AA{A\&A}
\def\JLTP{{ Journal of Low Temperature Physics}}
\def\SAL{{ Sov. Astron. Lett.}}

\def\be{\begin{equation}}
\def\ee{\end{equation}}
\def\bea{\begin{eqnarray}}
\def\eea{\end{eqnarray}}

\usepackage{amsmath}
\usepackage{amssymb}

\newcommand{\Photo}{\includegraphics[height=35mm]{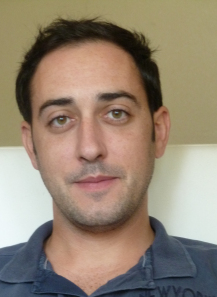}}

\begin{document}
%\vspace*{1.5cm}
\vspace*{4cm}
\title{COMPONENT SEPARATION FOR FUTURE CMB B-MODE SATELLITES}

\author{ M. REMAZEILLES }

\address{Jodrell Bank Centre for Astrophysics, School of Physics and Astronomy, Alan Turing Building, \\The University of Manchester,
Oxford Road, Manchester, M13 9PL, U.K.}

\maketitle\abstracts{Next-generation CMB satellite concepts (\emph{LiteBIRD}, \emph{CORE}, \emph{PIXIE}, \emph{PICO}) are being proposed to detect the primordial CMB B-mode polarization at large angular scales in the sky for tensor-to-scalar ratio values of ${r \lesssim 10^{-3}}$. Yet undetected, primordial CMB B-modes will provide the unique signature of the primordial gravitational waves of quantum origin predicted by inflation. We present recent forecasts on the detection of the primordial CMB B-modes in the presence of astrophysical foregrounds and gravitational lensing effects,  in the context of the proposed CMB space mission \emph{CORE}. We also discuss the problem of foregrounds and component separation for the search for primordial B-modes, and highlight specific challenges in this context: frequency range, spectral degeneracies, foreground modelling, spectral averaging effects.}

\section{Introduction}\label{sec:intro}

The search for the primordial B-mode polarization of the cosmic microwave background (CMB) radiation at large angular scales in the sky is one of the most exciting challenge of modern cosmology, because such a signal would be the direct signature of the primordial gravitational waves predicted by inflation~\cite{inf}. The amplitude of the primordial B-mode signal, termed as tensor-to-scalar ratio, $r$, will determine the energy scale of inflation, ${E_{\rm inf} \simeq (r/0.008)^{1\over 4}\,10^{16}}$\,GeV. CMB satellite concepts (\emph{LiteBIRD}~\cite{ltb}, \emph{CORE}~\cite{core}, \emph{PIXIE}~\cite{pixie}, \emph{PICO}~\cite{pico}) are being proposed to detect large-scale CMB B-modes at ${r \lesssim 10^{-3}}$. This is a real challenge because the signal is extremely faint ($\lesssim 50$\,nK r.m.s. fluctuations in the sky) and obscured by very bright polarized Galactic foreground emissions by many orders of magnitude. In addition, gravitational lensing effects by large-scale structures transform CMB E-modes into noise-like B-modes, while spurious B-modes are created by instrumental systematic effects. In this context, component separation methods are critical to subtract the foregrounds and extract the CMB B-mode signal, since the residual foreground contamination will set the ultimate uncertainty limit with which $r$ can be measured.

In this article, we report on recent B-mode detection forecasts with the CMB satellite concept \emph{CORE}~\cite{core2}, and briefly discuss about the problem of foregrounds and component separation for B-modes, by highlighting subtle issues that arise in this context.

\section{B-mode component separation forecasts for \emph{CORE}}\label{sec:core}

%We report on the results~\cite{core2} of component separation and primordial CMB B-mode reconstruction, based on sky simulations of the proposed space mission \emph{CORE}~\cite{core}, that is designed to observe the full sky through 19 frequency bands, ranging from $60$ to $600$\,GHz, with high sensitivity.
%\emph{CORE} is designed to observe the full sky through 19 frequency bands, ranging from $60$ to $600$\,GHz, with a combined sensitivity of $1.7$\,$\mu$K.arcmin in polarization and a resolution of $\lesssim 10$ arcmin at CMB frequencies~\cite{core}.  

The proposed space mission \emph{CORE}~\cite{core} is designed to observe the full sky with high sensitivity through 19 frequency bands, ranging from $60$ to $600$\,GHz. We report on the results~\cite{core2} of component separation and primordial CMB B-mode reconstruction, based on \emph{CORE} sky simulations.

\subsection{Sky simulations}\label{subsec:prod}

Using the {\sc PSM} (Planck Sky Model) software~\cite{psm}, we have simulated full-sky polarization maps for the $19$ frequency bands ($60$ to $600$ GHz) of \emph{CORE}. Our simulated sky maps~\cite{core2} include: CMB E- and B-mode polarization, with an optical depth to reionization $\tau=0.055$ and a tensor-to-scalar ratio ranging from $r=10^{-3}$ to $10^{-2}$; lensing E- and B-modes; Galactic and extra-galactic foreground polarization. Galactic foregrounds consist of thermal dust emission, based on the \emph{Planck} {\sc GNILC} dust template~\cite{gnilc} at $353$\,GHz, with average polarization fraction of $5$-$10$\% over the sky; polarized Galactic synchrotron emission, as observed by \emph{WMAP} at $23$\, GHz~\cite{mamd}; and Galactic anomalous microwave emission (AME) with 1\% polarization fraction. Extra-galactic foregrounds include compact radio and infrared sources with respectively $3$\%-$5$\% and $1$\% mean polarization fractions. The dust map is interpolated across the \emph{CORE} frequency bands through a modified blackbody (MBB) emission law having variable spectral index and temperature over the sky, with mean values $\langle\beta_d\rangle=1.6$ and $\langle T_d\rangle=19.4$\,K, as measured by \emph{Planck}~\cite{gnilc}. The synchrotron map is extrapolated across frequencies through a power-law with an average spectral index of ${\langle\beta_s\rangle=-3}$ varying over the sky~\cite{mamd}. The emission law for extrapolating the AME component is modelled by assuming a Cold Neutral Medium~\cite{ame}. Compact source templates are extrapolated across \emph{CORE} frequencies by assuming random steep or flat power-laws for radio sources, and both modified blackbodies and power-laws for infrared sources. The component maps at each frequency are coadded, convolved by a Gaussian beam using the \emph{CORE} FWHM values, and instrumental white noise is added to each frequency map using the sensitivities quoted by \emph{CORE}~\cite{core}.

\subsection{Component separation methods}\label{subsec:methods}

We have applied four independent component separation algorithms~\cite{core2} to the \emph{CORE} sky simulations to perform foreground removal, reconstruction of the CMB B-mode power spectrum, and estimation of the tensor-to-scalar ratio: {\sc Commander}~\cite{compsep}, a Bayesian parametric method for a multi-component pixel-by-pixel spectral fit using MCMC Gibbs sampling; {\sc Smica}~\cite{compsep}, a blind method for a power-spectra fit in harmonic space; {\sc Nilc}~\cite{compsep}, a blind method for minimum-variance internal linear combination in wavelet space; and {\sc xForecast}~\cite{xforecast}, an alternative parametric fitting approach in pixel space. The first three algorithms have already a strong heritage from real \emph{Planck} data analysis~\cite{compsep}. Parametric methods are only limited by the accuracy with which the foregrounds are modelled in the fit, while blind methods do not rely on any assumptions about the foregrounds but are limited by the overall variance of the foregrounds and the number of frequency channels and multipole modes available to minimize this variance. Since the variance of the foregrounds is much larger at the reionization scales ($\ell \simeq 10$), parametric fitting was preferred to reconstruct CMB B-modes at low multipoles $\ell < 50$ (reionization peak), while blind methods were used to reconstruct the signal at large multipoles $\ell \geq 50$ (recombination peak).

\subsection{Results}\label{subsec:results}

\begin{figure}%[htbp]
\vspace{-1cm}
\centering
    \includegraphics[width=0.4\linewidth]{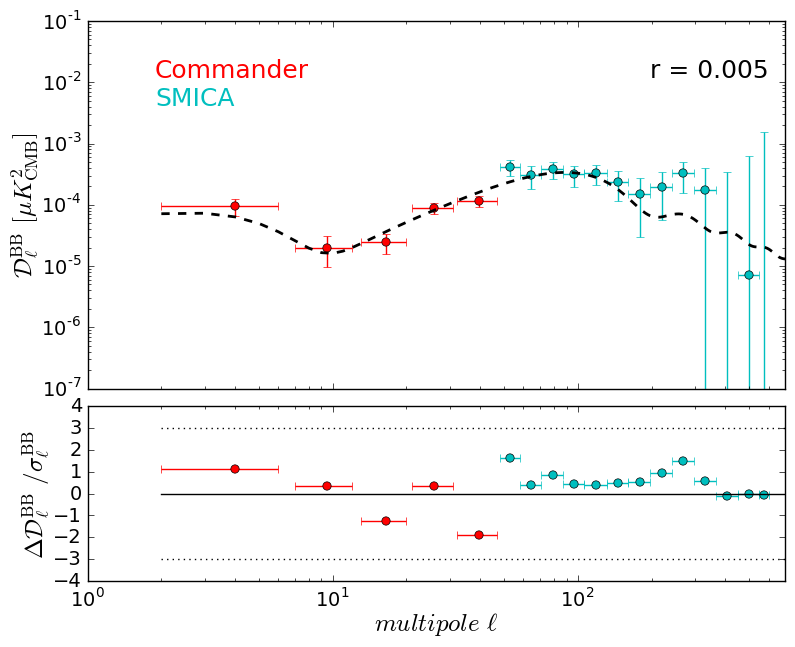}~%\hfill
    \includegraphics[width=0.43\linewidth]{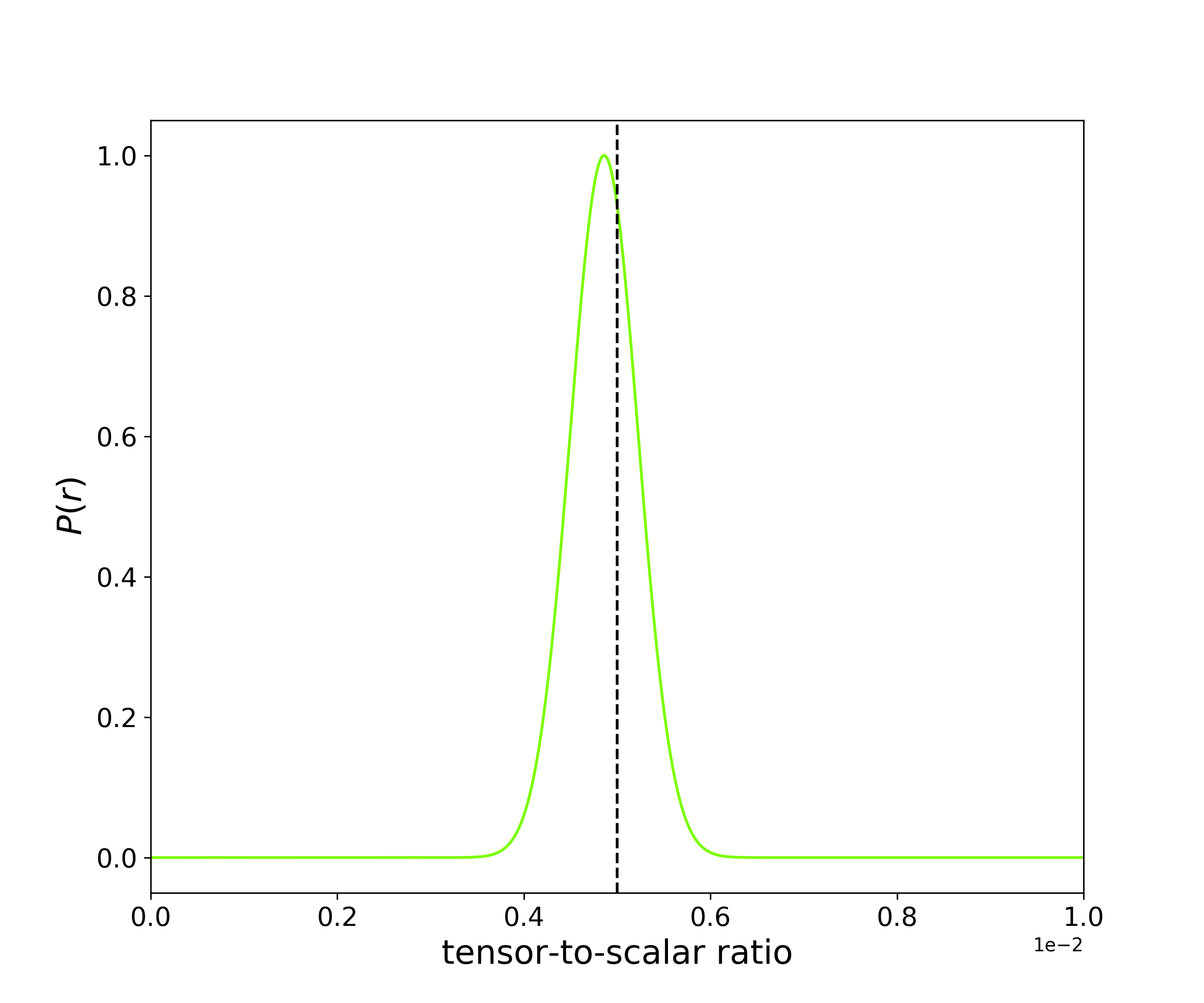}~\\
\caption[]{Primordial B-mode reconstruction at $r=5\times 10^{-3}$ (\emph{left}) and estimate of $r$ (\emph{right}) for \emph{CORE}.}
\label{fig:core}
\end{figure}

The left panel of Fig.~\ref{fig:core} shows the reconstruction of the primordial CMB B-mode after foreground cleaning with {\sc Commander} and {\sc Smica} for a fiducial tensor-to-scalar ratio of ${r = 5\times 10^{-3}}$, in the absence of lensing. The broad frequency range of \emph{CORE} allows us to recover the primordial B-mode signal at both reionization and recombination peaks, and to measure the posterior distribution of ${r = 5\times 10^{-3}}$ without bias at $12\sigma$ significance (right panel of Fig.~\ref{fig:core}) after foreground cleaning. In the presence of lensing contamination, a shortcut was adopted to perform delensing. Instead of correcting for the lensing variance in the foreground-cleaned CMB B-mode map, as real delensing approaches would do, we left $40$\% of the lensing B-mode power in the CMB map realization of the simulation, then performed foreground cleaning on the modified simulation. This is equivalent to performing foreground cleaning and $60$\% delensing, which is the delensing capability quoted by \emph{CORE}~\cite{core3}. 
%The result is shown on the right panel of Fig.~\comment{[Fig.]}. 
In the presence of lensing, ${r = 5\times 10^{-3}}$ is detected at $4\sigma$ significance after foreground cleaning and $60$\% delensing~\cite{core2}, putting \emph{CORE} in an excellent position to constrain the energy scale of inflation for the Starobinsky's $R^2$ inflation model~\cite{inf}.%, for which ${r = 4.2 \times 10^{-3}}$.

For a tensor-to-scalar ratio as low as $r=10^{-3}$, the residual foreground contamination in the CMB B-mode power spectrum after component separation is significant at all angular scales for all the methods~\cite{core2}, resulting in a $3\sigma$ bias on the measurement of $r=10^{-3}$ by \emph{CORE}. The bias is attributed to the available frequency range $60$-$600$ GHz of \emph{CORE}, for which the minimized variance of the foregrounds achieved by blind methods ({\sc Nilc} and {\sc Smica}) still exceeds $r=10^{-3}$ in power while being lower than $r=5\times 10^{-3}$. For parametric methods ({\sc Commander}), the absence of frequencies below $60$\,GHz prevent the synchrotron spectral index, $\beta_s$, to be constrained at the level of precision required for $r=10^{-3}$: while the recovered distribution of $\beta_s$ over the sky has same mean and standard deviation than the actual distribution, it is more Gaussian-distributed, which results in a $2$\% mismatch on $\beta_s$. %in some directions of the sky. 
This error on $\beta_s$ is large enough to cause an excess B-mode power at a level of $r\approx 2.5 \times 10^{-3}$ when extrapolating synchrotron B-modes to CMB frequencies~\cite{core2}. Subpercent precision on foreground spectral indices is thus required to measure $r=10^{-3}$ without bias, which can be achieved with broader frequency ranges (Sect.~\ref{sec:discussion}).

\section{Concluding remarks: subtle issues for B-mode component separation}\label{sec:discussion}

\begin{figure}%[htbp]
\vspace{-1cm}
\centering
    \includegraphics[width=0.4\linewidth]{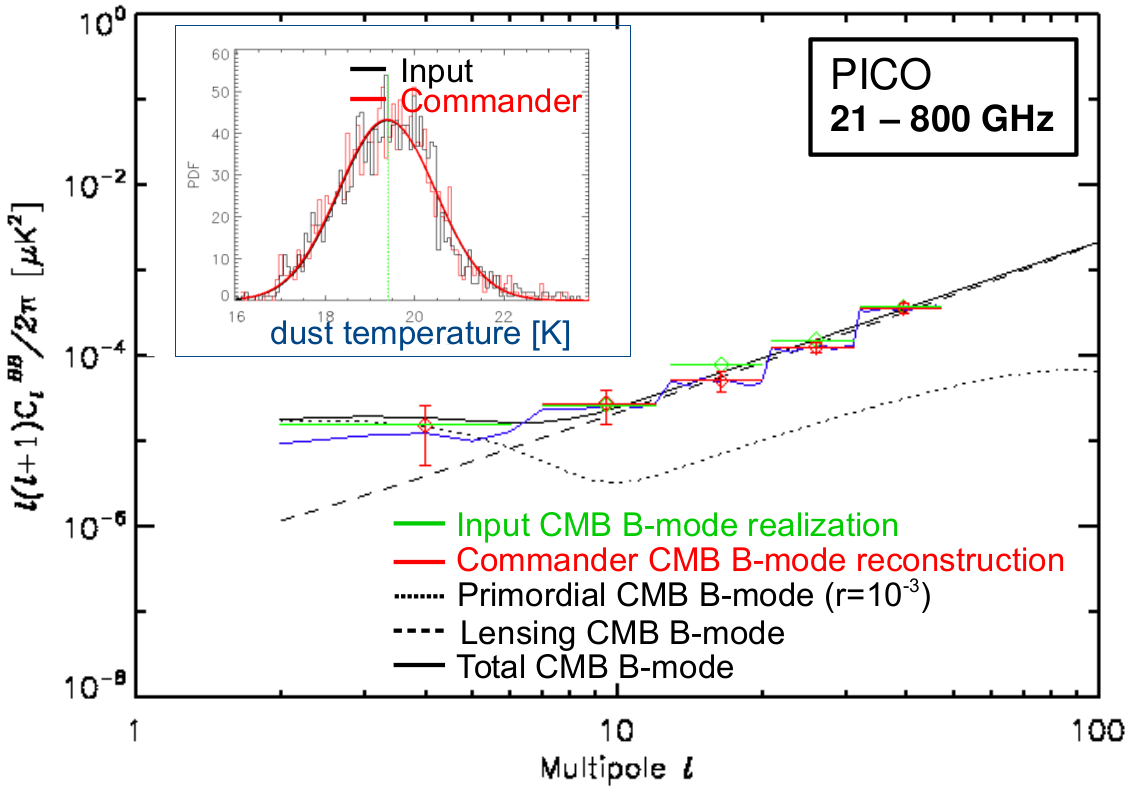}~%\hfill
    \includegraphics[width=0.4\linewidth]{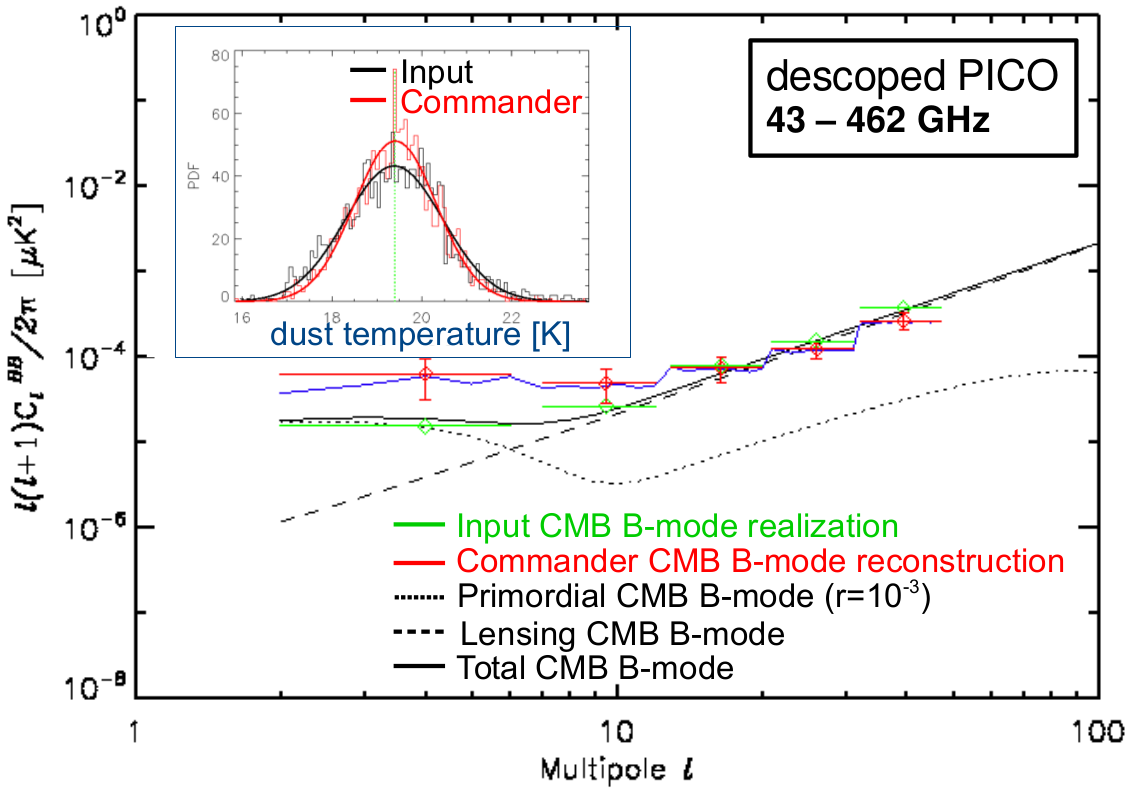}~\\
\caption[]{CMB B-mode reconstruction for \emph{PICO} 21-800 GHz (\emph{left}) versus \emph{descoped PICO} 43-462 GHz (\emph{right}).}
\label{fig:pico}
\end{figure}

{\bf On the importance of a broad frequency range.} The CMB satellite concept \emph{PICO}~\cite{pico} benefits from a broader frequency range ($21$-$800$\,GHz) than \emph{CORE}. The reconstruction of the CMB B-mode power spectrum at $r=10^{-3}$ with \emph{PICO} is shown in the left panel of Fig.~\ref{fig:pico}, for the same sky simulation. Due to a larger frequency range of $21$-$800$ GHz, \emph{PICO} allows {\sc Commander} to control the foreground contamination at the desired accuracy to measure ${r=10^{-3}}$ with $2.5\sigma$ significance, without any bias, from low multipoles $2\leq \ell \leq 50$. Conversely, narrowing the baseline frequency range of \emph{PICO} to $43$-$462$\,GHz (right panel of Fig.~\ref{fig:pico}) introduces a bias at large angular scales on the recovered B-mode power spectrum because of residual dust contamination. In the absence of high frequencies $\gtrsim 400$\,GHz, the dust MBB temperature is constrained with lower accuracy (left-corner stamp in the right panel of Fig.~\ref{fig:pico}), which results in spectral degeneracies in the fit and translates into a bias on the reconstructed CMB B-mode at $r=10^{-3}$.

%\noindent
{\bf Foreground mismodelling.} Due to the very large dynamic range between foregrounds and CMB B-mode fluctuations, component separation for polarization is much more sensitive to foreground modelling uncertainties than for temperature. Mismodelling two MBB dust components as a single MBB dust component in the {\sc Commander} fit was shown to bias $r=5\times 10^{-2}$ by more than $3\sigma$ for any CMB satellite concept~\cite{bias}. Most important, CMB experiments with narrower frequency ranges show no chi-square evidence for incorrect dust modelling~\cite{bias}, the fit of the overall sky emission being still accurate in narrow frequency ranges while it suffers from spectral degeneracies. Frequencies below $60$\,GHz and above $400$\,GHz are thus critical for CMB B-mode experiments to get chi-square evidence for incorrect foreground modelling and false detections of $r$. It could be argued that increasing the frequency range of observations will introduce additional foregrounds. However, Galactic foregrounds are not fully decorrelated across frequencies, so that the increase in foreground complexity (extra degrees of freedom) should be more than compensated by the increase of information (extra frequencies) for component separation.  

%\noindent
{\bf Spectral averaging effects.} Foreground spectral indices vary in the sky from line-of-sight to line-of-sight, but sky map observations are pixelized and do not have infinite resolution, so that different spectral indices are averaged within pixels or beams~\cite{moments}. The averaging of power-laws with different spectral indices in a pixel is no longer a power-law, instead it introduces spurious curvatures in the effective emission law across frequencies~\cite{core2}. Say otherwise, the effective emission laws of the foregrounds on the pixelized maps may differ from the real emission laws in the sky. 
%It has been shown~\cite{core2} that degrading the pixel resolution of sky maps creates a spurious curvature in the effective dust emission law. 
%Degrading the pixel resolution of a sky map creates spurious curvatures in the effective dust emission law~\cite{core2}. 
Averaging effects are critical for parametric fitting methods in the context of B-modes. Ignoring in the parametric fit a spurious dust curvature of $0.05$ caused by averaging effects results in a bias of $\Delta r \gtrsim 10^{-3}$ on the tensor-to-scalar ratio~\cite{core2}. To tackle this issue, moment-expansion approaches~\cite{moments}, rather than astrophysical model fitting, might provide an interesting avenue.%, but their efficiency still needs to be demonstrated on B-modes.   

%\section{Conclusion}\label{sec:conclusion}

%\vspace{-0.8mm}
%\begin{figure}
%\begin{minipage}{0.5\linewidth}
%\centerline{\includegraphics[width=0.7\linewidth]{final_commander-smica_r0-005.png}}
%\end{minipage}
%\hfill
%\begin{minipage}{0.5\linewidth}
%\centerline{\includegraphics[width=0.7\linewidth]{likelihood_r_model18v5_r5em3_hybrid_joint-smica.png}}
%\end{minipage}
%\caption[]{same figure with draft option (left), normal (center) and rotated (right)}
%\label{fig:radish}
%\end{figure}

\section*{Acknowledgments}

The author acknowledges funding from the ERC Consolidator Grant {\it CMBSPEC} (No.~725456).

\end{document}